\documentclass[aps,prb,twocolumn,superscriptaddress]{revtex4-1}
\usepackage{hyperref}
\usepackage{dsfont}
\usepackage{amsmath}
\usepackage{amssymb}
\usepackage{graphicx}
\usepackage{color}

\newcommand{\diff}{\mathrm{d}}  
\newcommand{\e}{\mathrm{e}} 
\newcommand{\sign}{\operatorname{sign}}
\newcommand{\Ham}{\mathcal{H}}  

\newcommand{\xp}{x^{\prime}}  

\newcommand{\w}{\omega}    

\newcommand{\pf}{p_{\mathrm{F}}}  
\newcommand{\ef}{\epsilon_{\mathrm{F}}}  
\newcommand{\vf}{v_{\mathrm{F}}}  

\newcommand{\nB}{n_{\mathrm{B}}}  

\begin{document}


\title{Plasmon localization and relaxation, and thermal transport in one-dimensional conductors}


\author{M. Bard}
\affiliation{Institut f\"ur Nanotechnologie, Karlsruhe Institute of Technology, 76021 Karlsruhe, Germany}
\author{I.~V. Protopopov}
\affiliation{Department of Theoretical Physics, University of Geneva, 1211 Geneva, Switzerland  }
\affiliation{Landau Institute for Theoretical Physics, 119334 Moscow, Russia}
\author{A.~D. Mirlin}
\affiliation{Institut f\"ur Nanotechnologie, Karlsruhe Institute of Technology, 76021 Karlsruhe, Germany}
\affiliation{Institut f\"ur Theorie der Kondensierten Materie, Karlsruhe Institute of Technology, 76128 Karlsruhe, Germany} 
\affiliation{Petersburg Nuclear Physics Institute, 188350 St. Petersburg, Russia}
\affiliation{Landau Institute for Theoretical Physics, 119334 Moscow, Russia}


\date{\today}

\begin{abstract}
We study the localization and decay properties as well as the thermal conductance of one-dimensional plasmons. Our model contains a Luttinger-liquid part with spatially random plasmon velocity and interaction parameter as well as a nonlinearity that is cubic in density. The scaling of the decay rate of plasmons is obtained in several regimes. 
At sufficiently high frequencies, it describes the inelastic life time of localized plasmon excitations that crosses over to the clean result with lowering frequencies. 
For higher frequencies, we analyze implications of many-body-localization effects that lead to a suppression of the decay rate. We find that the thermal conductance depends  in a non-trivial fashion on the system size $L$. Specifically, it scales as $L^{-1/2}$ for sufficiently short wires and crosses over to $L^{-2/3}$ scaling for longer wires. 
\end{abstract}


\maketitle


\section{Introduction}
\label{Sec:Intro}

One-dimensional (1D) quantum systems show a plenty of fascinating phenomena \cite{GiamarchiBook}. Interaction effects are often strong in 1D, so that the proper  elementary  excitations of the systems  are plasmons (collective density fluctuations) and not the genuine constituent  particles (fermions or bosons).
The resulting state of matter is referred to as the Luttinger liquid (LL). Due to advances in technological fabrication processes, many physical realizations of quantum interacting 1D systems are experimentally available. 

A research direction that has attracted a considerable amount of interest in recent years is the thermal transport and energy relaxation in various 1D setups, including systems of photons \cite{MeschkeEtAl06} and of cold atoms \cite{Esslinger2013,Esslinger2018} as well as quantum Hall edges with counterpropagating modes \cite{PierreEtAl10,PierreEtAl12,YacobyEtAl12,PierreEtAl13,HeiblumEtAl14,PierreEtAl17,HeiblumEtAl17,HeiblumEtAl17_anyonic,HeiblumEtAl18}. The quantum Hall structures
provide a particularly suitable  experimental platform for the  exploration of  this class of phenomena. Recent realizations of synthetic  quantum Hall edges \cite{KhrapaiEtAl14, Cohen2019}  allow for unprecedented  control of the system parameters.

Another prominent example of quantum 1D systems is the Josephson junction  (JJ) chains. They  were studied extensively during the last two decades in the
context of the superconductor-insulator transition 
\cite{Chow1998, HavilandEtAl00, Kuo2001, CedergenEtAl17} and as a platform for quantum computations\cite{Manucharyan2009_1,Manucharyan2009_2,Nguyen2018, Vool2014} and metrology\cite{GuichardHekking10}.
The relaxation of plasmonic waves in JJ chains was probed recently via  spectroscopic measurements\cite{ManucharyanEtAl18}, which has triggered intensive theoretical studies \cite{BardEtAl18,WuSau18,HouzetGlazman19} of the mechanisms responsible for the broadening of bosonic states in these systems in various parameter regimes.

Disorder is inevitable in physical systems and is particularly important in low dimensions. 
It induces the phenomenon of Anderson localization\cite{abrahams201050} that can have a dramatic impact on transport properties. An interplay of disorder and interaction is in general a difficult problem, with various facets. In particular, the interaction-induced inelastic processes tend to establish decoherence of excitations, thus reducing the effect of Anderson localization that is crucially based on quantum coherence. It has been realized, however, that this delocalizing effect of interaction may be suppressed (strongly or even completely) due to quantum localization in the many-body space. This has has opened a new research area of many-body localization (MBL) \cite{Mirlin2005, Basko2006}, see
Refs. \onlinecite{Nandkishore2015,AletRev2018,AbaninRev2019} for recent reviews. The full MBL implies non-ergodicity, i.e., the complete breakdown of the conventional statistical mechanics.

The precise nature of disorder and its physical consequences may vary. 
In a generic 1D conductor, impurities cause  backscattering of charge carriers. Their interplay with interactions is highly non-trivial\cite{GiamarchiSchulz88,GMP07}. On the one hand, interactions strongly renormalize disorder seen by low-energy electrons. On the other hand, the real (as opposed to virtual) scattering processes enabled by the interactions pave the way to energy relaxation and dephasing. The interaction-induced dephasing  then cuts off the singular Anderson localization corrections and is responsible for the finite electric conductivity of the system at not too strong disorder and moderate temperatures.

In the context of the JJ chain, the disorder that is commonly considered  are the random stray charges.  Their relevance to the properties of the system  stems from the fact that they can couple directly (via the Aharonov-Casher effect) to the  quantum phase slips\cite{BradleyDoniach84,MatveevEtAl02} (QPS) that are the driving force of the  quantum superconductor-insulator transition.   

A disorder of different kind can also be present and, under special circumstances, play the major role in the aforementioned systems. Specifically, if the impurity potential experienced  by the electrons in a quantum wire is smooth on the scale of the Fermi wavelength, the associated backscattering is strongly suppressed and can be ignored in a wide range of parameters.  On the other hand, the induced inhomogeneities of the average electronic density cause the spatial variations  of effective interaction between particles.  Those variations translate then   into the inhomogeneities of the Luttinger parameter entering the LL description of the system and lead to backscattering (and, as a result, to localization) of {\it plasmons}\cite{GramadaRaikh97}. While the low-frequency charge transport    is  unaffected by the plasmon backscattering, the latter  has a profound impact on the energy propagation in the system \cite{KhmelnitskiiEtAl98,FilipponeEtAl16} causing in particular strong violation of the celebrated Wiedemann-Franz law. 

In a similar manner, in JJ chains with sufficiently strong Josephson couplings, the QPS are strongly (exponentially) suppressed. The system is then superconducting from the point of view of the charge transport (at least, up to exponentially low temperatures and exponentially large lengths that are often beyond the practical reach). Yet, the local fluctuations of the parameters of the system (Josephson and charging energies) create disorder for plasmonic waves and alter their dynamics.

Beyond the harmonic approximation, the plasmons in the  aforementioned physical setups typically interact via a local cubic interaction. For example, in  quantum wires,  such an interaction originates from 
the quadratic curvature of the dispersion relation of the genuine electrons\cite{Schick68,Jevicki1980,KhmelnitskiiEtAl98}.   In the clean case,  the effects of the
plasmonic 
interactions on the dynamical correlation functions were intensively  discussed in recent years, leading to the emergence of the concept of non-linear Luttinger liquids \cite{Review_Glazman12}. In particular, it was shown recently\cite{SamantaEtAl19} that the cubic interaction of plasmons governs  the  thermal conductance of a clean LL at lowest temperatures. The resulting thermal conductance displays a non-trivial scaling with the size of the system, $G(L)\propto L^{-2/3}$, the behavior also known\cite{Narayan02,Spohn14} to occur in classical translationally invariant non-linear 1D systems.  

In this paper we investigate a LL with a spatially random Luttinger parameter and cubic interaction of bosons. Our work is largely motivated by experimental developments and prospects discussed above. The model that we consider is expected to be relevant to various realizations of correlated 1D systems. 
 We explore the interaction-induced decay of the plasmons in its dependence on the frequency and temperature.  We find that at lowest frequencies the disorder plays a minor role, and the corresponding decay rate   coincides with the one found\cite{Samokhin98} previously in a clean LL,  $1/\tau\propto  \w^{3/2}$. At higher frequencies, the disorder effects become crucially important. The relaxation rate  grows linearly with frequency and eventually saturates due to a weak form of MBL effects. 
 Further, joint effect of disorder and plasmonic anharmonicity manifests itself  in the behavior of the thermal conductance as a function of the system length $L$.
 In a short system the thermal conductance is fully controlled by disorder, yielding $G\propto L^{-1/2}$. Upon increase of $L$, the system crosses over to the regime where $G$ is governed by an interplay of disorder and interaction, $G\propto L^{-2/3}$.


The paper is organized as follows. In Sec.~\ref{Sec:Model} we give a precise definition of our model. A brief discussion of the localization of the  non-interacting plasmons is presented in Sec.~\ref{Sec:Localization_Length}.  In  Sec.~\ref{Sec:lifetime}  we  investigate  the interaction-induced  lifetimes of plasmons in various regimes.  Section \ref{Sec:thermal_conductance} is devoted to the analysis of the thermal conductance. Finally, Sec.~\ref{Sec:Summary} contains a summary of our results and a discussion of prospective research directions.



\section{Model}
\label{Sec:Model}

The Hamiltonian of our model describing plasmons in a disordered one-dimensional system consists of two parts,
\begin{equation}
\Ham=\Ham_0+\Ham_1.
\label{Eq:Hamiltonian}
\end{equation}
The quadratic part
\begin{equation}
\Ham_0=\frac{1}{2\pi}\int \diff x \left[v(x) K(x) \left(\partial_x\theta\right)^2+\frac{v(x)}{K(x)}(\partial_x \phi)^2\right]
\label{Eq:H0}
\end{equation}
is a generic Luttinger-liquid Hamiltonian with space-dependent plasmon velocity $v(x)=v_0+\delta v(x)$ and Luttinger parameter $K(x)=K_0+\delta K(x)$. The random fluctuations $\delta v(x)$ and $\delta K(x)$ are assumed to have zero mean and Gaussian statistics with
\begin{equation}
\begin{split}
\overline{ \delta K(x) \delta K(\xp)}&=D_K \delta(x-\xp),
\\
\overline{\delta v(x) \delta v(\xp)}&=D_v \delta(x-\xp),
\\
\overline{ \delta v(x) \delta K(\xp)}&=D_{vK} \delta(x-\xp).
\end{split}
\label{Eq:correlations}
\end{equation}
The fields $\partial_x \phi$ and $\partial_x\theta$ are related to  the  particle density  and current, respectively,
\begin{equation}
\partial_x\phi=-\pi \rho, \qquad v K \partial_x\theta=\pi j.
\end{equation}
They satisfy the commutation relations
\begin{equation}
[\phi(x),\partial_{\xp}\theta(\xp)]=i\pi \delta(x-\xp).
\end{equation} 

One particular  instance of such a model is a quantum wire with smooth (on the scale of the Fermi wavelength) disorder considered in Refs. 
\onlinecite{GramadaRaikh97,KhmelnitskiiEtAl98}, see also appendix~\ref{App:comparision-Khmelnitskii}.  Here, the smoothness of the  disorder suppresses backscattering of electrons, whereas the spatial  variations of the equilibrium fermionic density translate into fluctuations of the velocity $v$ and the effective interaction~$K$.  

 Another physical realization of the Hamiltonian (\ref{Eq:H0}) is the low-energy limit of  a JJ chain with sufficiently strong Josephson couplings precluding quantum phase slips\cite{BradleyDoniach84,MatveevEtAl02}. In this case, the variances $D_K$, $D_v$ and $D_{Kv}$ can be expressed in terms of the 
disorder in local Josephson and charging energies. 

The interaction effects are described in our model by the cubic coupling of bosons 
\begin{equation}
\Ham_1=-\frac{1}{6\pi m}\int \diff x \left[3(\partial_x \theta)^2 \partial_x\phi+(\partial_x\phi)^3\right].
\label{Eq:H1}
\end{equation}
The origin of this interaction term  is particularly transparent in the case of a quantum wire. Equation (\ref{Eq:H1}) expresses in the bosonization language   the quadratic correction,  $\delta\epsilon(k)=k^2/2m$, to the spectrum of the constituent fermions near the Fermi points \cite{Schick68,Jevicki1980}. It can be traced back to the dependence of the ground state energy density of   fermions with mass\footnote{The precise meaning and the value of the fermionic mass $m$ in Eq. (\ref{Eq:H1}) may depend on the short-distance cutoff imposed on the Hamiltonian (\ref{Eq:Hamiltonian}). It coincide with the bare electron mass, if the cutoff of the order of Fermi momentum is assumed, or may include some interaction-induced renormalization if the theory is regularized at smaller momenta. } $m$ on the particle density and macroscopic velocity $V$, $E(\rho, V)=\pi^2\rho^3/6m+m \rho V^2/2$.

If the Hamiltonian (\ref{Eq:Hamiltonian}) is viewed as an effective  low-energy field theory of a JJ chain,  the anharmonicity (\ref{Eq:H1}) is also natural.  Indeed, any perturbation allowed by symmetry  is expected to appear in such an effective description, and the two terms in the square brackets in Eq. (\ref{Eq:H1}) are precisely  the two possible perturbations of the lowest scaling dimension  that can be added to the fixed-point LL Hamiltonian.  Unlike the case of fermions with quadratic spectrum where the Galilean  
invariance enforces the equality of the coefficients in front of the aforementioned terms, no such connection exists generically in the context of JJ chains. This distinction in the description of the two systems  is, however, irrelevant for us as it can only alter numerical coefficients in our results.  As to the microscopic value of the ``mass'' $m$ 
(or, more generally, of the two allowed coupling constants) in the case of JJ chains, it can be related to the ground state energy of the system, cf. discussion after Eq. (\ref{Eq:H1}) as well as Refs. \onlinecite{Pereira2006, Lukyanov1998} where a similar problem was discussed in the context of the XXZ spin model. The cubic anharmonicity (\ref{Eq:H1}) is thus expected to be a perturbation generically present in systems described by LL theory. In this work we treat the mass $m$ as a model parameter, leaving aside  the question of its microscopic description. 

Let us emphasize that, throughout this work, we limit our consideration to the case of weak uncorrelated disorder. The strong-disorder effects (e.g., the situation of a heavy-tailed  distribution of the Josephson couplings  in a JJ chain\cite{Refael2013}) or the effects of correlated disorder (arising, e.g., in the description of plasmonic modes existing on top of a pinned charge-density wave\cite{Gurarie2002, Gurarie2003, HouzetGlazman19})  may alter significantly\cite{Gurarie2002, Gurarie2003, Refael2013,  Amir_Oreg_Imry18} the density of states and the properties of wave functions of low-energy plasmons. Their interplay with the anharmonicities in the system constitute an interesting direction for future research but is beyond the scope of the present work.



\section{Localization length of noninteracting plasmons\label{Sec:Localization_Length}}

In the absence of the anharmonic term \eqref{Eq:H1}, the plasmons do not interact and get localized due to the disorder, with the localization length equal to the 
mean free path with respect to backward scattering. In order  to evaluate the latter,  we decompose the Hamiltonian into the homogeneous and disorder  parts, 
$
\Ham_0=\Ham_0^{\mathrm{LL}}+\Ham_0^{\mathrm{dis}}
$, 
 given by
\begin{eqnarray}
\Ham_0^{\mathrm{LL}}=\frac{1}{2\pi}\int \diff x \left[v_0 K_0 \left(\partial_x \theta\right)^2+\frac{v_0}{K_0}\left(\partial_x \phi\right)^2\right]
\label{Eq:homogeneous_Hamiltonian}
\end{eqnarray}
and
\begin{eqnarray}
\Ham_0^{\mathrm{dis}}&=& \frac{1}{2\pi}\int \diff x \,\delta v(x) \left[K_0 \left(\partial_x\theta\right)^2+\frac{1}{K_0}\left(\partial_x \phi\right)^2\right] \nonumber
\\
&+& \frac{v_0}{2\pi}\int \diff x \, \delta K(x)\left[\left(\partial_x\theta\right)^2-\frac{1}{K_0^2}\left(\partial_x\phi\right)^2\right].
\label{Eq:perturbation}
\end{eqnarray}
We then compute the transport  scattering rate  for plasmons, treating the disorder perturbatively, and find   (see  appendix \ref{App:localization-length} for details)
\begin{equation}
\xi(\w)=\frac{v_0^2K_0^2}{2 D_K \w^2}.
\label{Eq:localization-length}
\end{equation}
Note that the  fluctuations of velocity can only contribute to 
forward scattering  and the localization length is determined solely by $D_K$. The result (\ref{Eq:localization-length}) agrees with the one obtained previously  in Refs.~\onlinecite{GramadaRaikh97,KhmelnitskiiEtAl98} for a less general model of disordered Luttinger liquid. 



\section{Decay of plasmonic states \label{Sec:lifetime}}

 In this section we explore the impact of the nonlinear term \eqref{Eq:H1} that  gives rise to finite lifetime of the localized plasmonic states discussed in Sec. \ref{Sec:Localization_Length}. Throughout this section we assume that the disorder is weak from the point of view of plasmons  involved in the scattering processes:
  the localization length of each plasmonic state is much longer than the corresponding de Broglie wavelength.  This condition 
is automatically fulfilled for all plasmons relevant to the transport phenomena (i.e, those with the energy smaller than temperature)  if the temperature is low enough,
\begin{equation}
T<\w_{\ast}\equiv\frac{v_0 K_0^2}{D_K}.
\label{Eq:w_star}
\end{equation}

The structure of this section is as follows. 
In Sec.~\ref{Sec:golden-rule}, we compute the decay rate of localized plasmons perturbatively by using the Fermi golden rule. We then analyze the limits of applicability of this golden-rule calculation and show that it is applicable within a (generically) broad rage of frequencies [determined by  Eqs. (\ref{Eq:w_tilde}) and  (\ref{Eq:w-bar})]  but fails both at sufficiently  low and sufficiently high frequencies.  The inelastic scattering of plasmons in those two regimes is discussed in Secs.~\ref{Sec:self-consistency} and~\ref{Sec:level-spacings}, respectively. 
In Sec. \ref{Sec:results} we summarize the results of this section, focusing on the sub-thermal plasmons relevant for the energy transport,  and analyze the mechanism of the motion of these plasmons through the system. 


\subsection{Golden-rule analysis\label{Sec:golden-rule}}

\begin{figure}
\center
\includegraphics[scale=0.3]{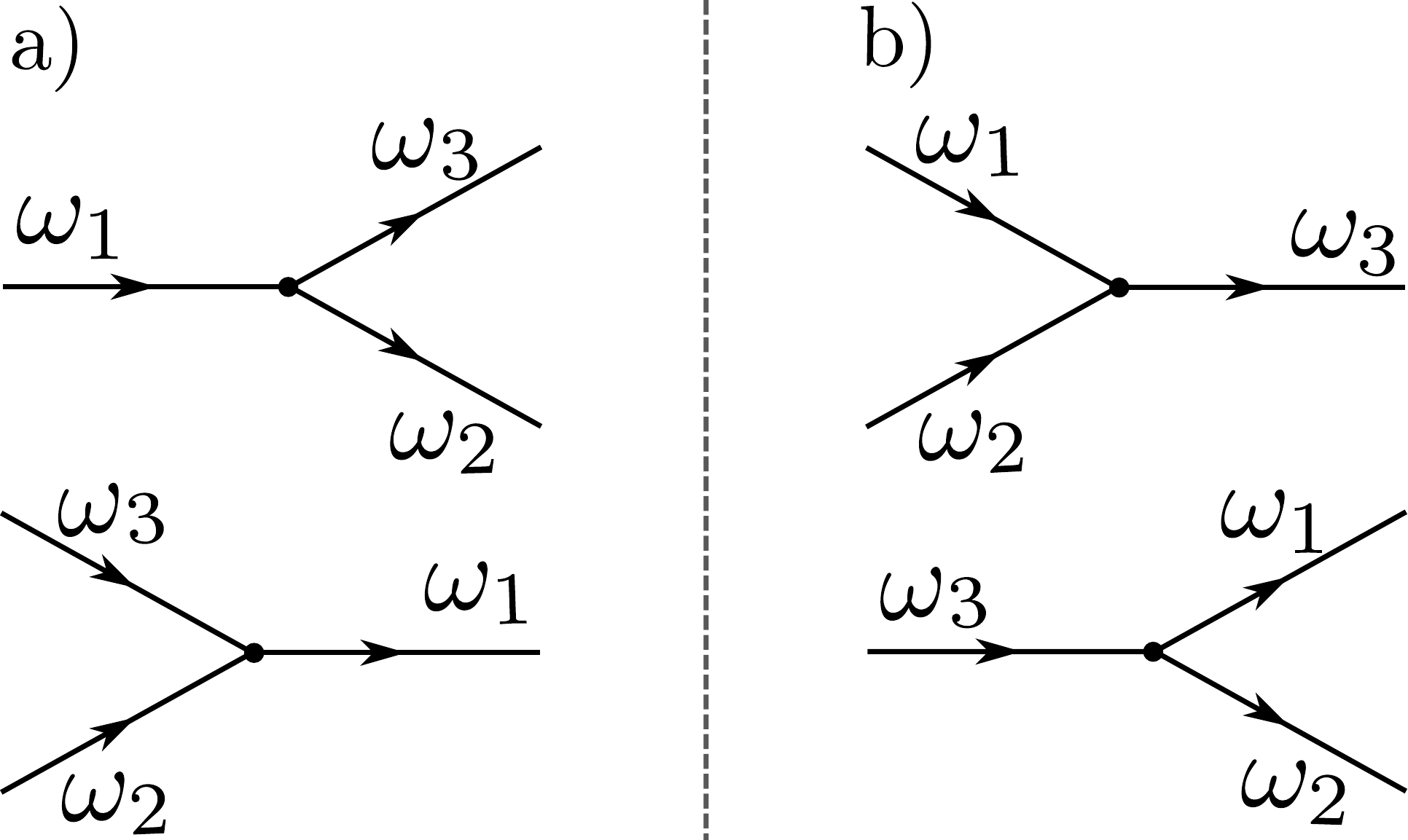}
\caption{Inelastic relaxation of a plasmonic mode at frequency $\omega_1$ via a) decay into two plasmons; b) collision with another plasmon. \label{Fig:processes}}
\end{figure}

In a clean Luttinger liquid,  the perturbation theory in the cubic interaction of bosons is highly singular. If the linear bosonic spectrum is assumed, the cubic non-linearity induces 
the one-into-two bosonic decay processes of the type shown in Fig. \ref{Fig:processes}. Within the Fermi golden-rule approximation,  the corresponding rate diverges  due to the equivalence of the energy and momentum conservation.  On the other hand, an arbitrarily small bending of the plasmonic spectrum makes it impossible to satisfy simultaneously the energy and momentum conservation in the one-into-two decay within the perturbative calculation\cite{Lin2013}. This singularity of the golden-rule analysis in a clean system requires a more sophisticated self-consistent analysis. We will return to this issue in Sec. \ref{Sec:self-consistency}.

In our system, the presence of disorder breaks  the momentum conservation, which  makes the lowest-order golden-rule calculation well defined (i.e., not singular).
Such a calculation was earlier performed in Ref. \onlinecite{KhmelnitskiiEtAl98} where the condition $\omega \gg T$ (opposite to the regime of our main interest in this work) was implicitly assumed.   We proceed now by performing the golden-rule calculation of the decay rate of localized plasmons in various ranges of frequency. After this, we will analyze the actual applicability of the golden-rule results.

We start from non-interacting plasmonic modes localized by disorder with the localization length  given by Eq.~\eqref{Eq:localization-length}. The quadratic Hamiltonian \eqref{Eq:H0} can be  diagonalized by a linear transformation of bosonic fields
\begin{equation}
\begin{split}
\phi(x)&=\sum_{\mu}\sqrt{\frac{\pi v(x) K(x)}{2\Omega_{\mu}}}\left(\psi_{\mu}(x)b_{\mu}+\psi_{\mu}^{\ast}(x)b_{\mu}^{\dagger}\right),
\\
\partial_x\theta(x)&=\frac{1}{i}\sum_{\mu}\sqrt{\frac{\pi \Omega_{\mu}}{2 v(x) K(x)}}\left(\psi_{\mu}(x)b_{\mu}-\psi_{\mu}^{\ast}(x)b_{\mu}^{\dagger}\right),
\end{split}
\label{Eq:transformation-to-bosons}
\end{equation}
where $\psi_{\mu}(x)$ are the eigenfunctions of the operator $\mathcal{D}$ defined by
\begin{equation}
\begin{split}
\mathcal{D}\psi_{\mu}&=-\sqrt{v(x)K(x)}\partial_x\left[\frac{v(x)}{K(x)}\partial_x\left(\sqrt{v(x)K(x)}\psi_{\mu}(x)\right)\right]
\\
&=\Omega_{\mu}^2\psi_{\mu},
\end{split}
\label{Eq:operator-D}
\end{equation}
and $b_{\mu}$, $b_{\mu}^\dagger$ are usual bosonic operators. In this basis, the quadratic Hamiltonian takes the form
\begin{equation}
\Ham_0=\sum_{\mu}\Omega_{\mu}\left[b_{\mu}^{\dagger}b_{\mu}+\frac{1}{2}\right].
\end{equation}
The eigenfrequencies of plasmonic states $\Omega_{\mu}$ are random and possess  Debye  (constant in 1D) density of states at low energy (cf. discussion at the end of Sec. \ref{Sec:Model}).

Under the assumption of weak disorder,  $|q| \xi \gg 1$, we model the wave functions   $\psi_{\mu}(x)$  by
\begin{equation}
\psi_{\mu}(x)=\frac{1}{\sqrt{\xi(\w)}}\e^{-|x-x_0|/\xi(\w)}\e^{i q x}, \quad \w=v_0 |q|, 
\label{Eq:psi}
\end{equation}
i.e. as plane waves modulated by an exponential factor reflecting the disorder-induced localization.  The   index $\mu$ in Eq. (\ref{Eq:psi}) comprises two parameters: the energy $\w$ and the center of the localized state $x_0$.
While the precise form of the wave functions in a disordered media can be intricate, Eq. (\ref{Eq:psi}) captures correctly the uncertainty in momentum of the plasmon and the associated broadening of the $\delta$-function expressing the momentum conservation in the scattering process.  This is sufficient for our analysis; we do not attempt to calculate numerical prefactors of order unity. 

 The inelastic scattering processes contributing to the relaxation of a bosonic mode at frequency $\omega_1$ are shown in~Fig.~\ref{Fig:processes}.
At temperatures satisfying the inequality  \eqref{Eq:w_star} (regardless of the relation between $\omega_1$ and $T$) 
the processes shown in Fig. \ref{Fig:processes}~b) contribute at most the same order of magnitude as the one-into-two plasmon decay shown in  Fig.~\ref{Fig:processes}~a).
We thus concentrate on the latter.  
The corresponding contribution to the decay rate can be extracted from the linearized bosonic kinetic equation treated in the relaxation time approximation and reads
\begin{equation}
\begin{split}
\frac{1}{\tau}=2\pi \int\frac{\diff x_2}{L}\int\frac{\diff x_3}{L}&\sum_{q_2,q_3}|M_{\mu_1}^{\mu_2,\mu_3}|^2\delta(\w_1-\w_2-\w_3)
\\
&\times[1+\nB(\w_2)+\nB(\w_3)],
\end{split}
\label{Eq:rate-initial}
\end{equation}
where $\nB(\w)$ is the equilibrium Bose distribution function and the matrix element for the process is given by
\begin{equation}
M_{\mu_1}^{\mu_2,\mu_3}=\langle 0|b_{\mu_3}b_{\mu_2}\Ham_1 b_{\mu_1}^{\dagger}|0\rangle, \quad \mu_i=(x_i,\w(q_i)).
\label{Eq:definition-matrix-element}
\end{equation}
Here the state $\mu_i$  (with $i=1,2,3$) is centered in the real and momentum space around  $x_i$ and $q_i$, respectively. 
 
Employing now  Eq. (\ref{Eq:psi}) and estimating the resulting integrals (see appendix \ref{App:decay-rate} for details) we find 
that the high-energy plasmons with $\omega>T$ relax at the rate
\begin{equation}
\frac{1}{\tau(\w)}\sim\frac{(3+K_0^2)^2\xi(\w)}{K_0 m^2 v_0^5}\w^4, \qquad \w > T.
\label{Eq:decay-rate-low-T}
\end{equation}
Here and subsequently  the symbol ``$\sim$'' stands for the  equality up to a factor of order unity (evaluation of which requires using exact statistics of wave functions of the localized plasmons).  The result (\ref{Eq:decay-rate-low-T}) agrees with the one found previously in Ref.~\onlinecite{KhmelnitskiiEtAl98}  [see Eq.~(12) of that work].

Within the linear-response regime, the thermal transport  is controlled by sub-thermal plasmons, $\omega < T$.  In this regime, we find the golden-rule decay rate
\begin{equation}
\frac{1}{\tau(\w)}\sim\frac{(3+K_0^2)^2\xi(\w)}{K_0m^2 v_0^5}\w^3 T, \qquad \w < T.
\label{Eq:decay-rate-high-T}
\end{equation}
Equation (\ref{Eq:decay-rate-high-T})  will be used below in the analysis of the thermal conductance. 

We turn now to the discussion of the applicability  conditions of the golden-rule calculation. First of all, the picture of localized plasmonic states decaying due to anharmonicities is only valid if the elastic scattering  time, $\tau_{\rm el}(\omega)\equiv \xi(\omega)/v$  is shorter then the inelastic time~(\ref{Eq:decay-rate-high-T}). 
The comparison of Eqs. (\ref{Eq:localization-length}) and (\ref{Eq:decay-rate-high-T}) (we focus here on the sub-thermal plasmons relevant for transport) reveals that this only holds for not too low frequencies,
\begin{equation}
\omega>\tilde{\w}\equiv\frac{K_0^3(3+K_0^2)^2T}{m^2 v_0^2 D_K^2}.
\label{Eq:w_tilde}
\end{equation}
For plasmons with  lower frequencies $\omega < \tilde{\w}$,  the inelastic time is shorter than the elastic one. In view of this, the disorder can be ignored (in the leading approximation) when one evaluates the rate of inelastic collisions for such low frequencies. We will analyze the inelastic relaxation of plasmons in this regime in Sec. \ref{Sec:self-consistency}.

Remarkably, the naive golden-rule approach fails not only at low $\omega$ but also at high enough frequencies. Specifically, the golden-rule  treatment is justified provided that the obtained rate is larger than the level spacing of final states into which a plasmon decays, see the related discussion in the context of MBL in quantum dots\cite{AGKL1997,Mirlin1997} and in strongly-disordered 1D electronic conductors\cite{Mirlin2005, Basko2006}.  In particular, for the  decay processes of Fig. \ref{Fig:processes}, the final states are the two-particle states (they are characterized by two frequencies $\omega_2$ and $\omega_3$ specifying the plasmonic modes that change their occupation as a result of the relaxation of the mode $\omega_1$). 
The golden rule approach breaks down at high frequencies because the localization of plasmons becomes stronger there and the two-particle level spacing grows with frequency. 
In  Sec. \ref{Sec:level-spacings} we will compute the associated level spacing and show that the  result (\ref{Eq:decay-rate-high-T}) remains valid for 
\begin{equation}
\omega<\bar{\w}\equiv\left(\frac{K_0^7(3+K_0^2)^2T}{D_K^4 m^2}\right)^{1/3}.
\label{Eq:w-bar}
\end{equation}
The plasmons with higher energies can not decay via the interaction with plasmons of similar energy and are in this sense in a  ``quasi MBL'' regime. The true MBL is, however, absent  because of the possibility to relax via the interaction with the ``bath'' of low-energy plasmons with frequencies $\omega<\bar{\w}$, see Sec. \ref{Sec:level-spacings}.


\subsection{Low frequencies: From disordered to clean regime\label{Sec:self-consistency}}

As discussed in Sec.~\ref{Sec:golden-rule}, our perturbative treatment of	the decay rate essentially relies on the broadening of plasmonic energy levels by disorder and   breaks down at low frequencies, $\w<\tilde{\w}$. In this low-frequency regime, the localization length is longer than the inelastic mean free path, so that disorder is essentially of no importance for the inelastic collision rate. 

We have also mentioned previously that  the perturbative treatment of the plasmonic decay in a clean LL is singular. The problem can be  circumvented by a {\it self-consistent} treatment of the  one-into-two decay channel \cite{Samokhin98}.
The result of Ref. \onlinecite{Samokhin98} translated to our notations reads\footnote{An alternative derivation of the characteristic scale (\ref{Eq:Samokhin}) was presented in Ref. \onlinecite{Arzamasovs2014}}
\begin{equation}
\frac{1}{\tau(\w)}\sim\frac{(3+K_0^2)T^{1/2} \w^{3/2}}{\sqrt{K_0}m v_0^2}.
\label{Eq:Samokhin}
\end{equation} 
It matches smoothly the rate (\ref{Eq:decay-rate-high-T}) at $\omega=\tilde{\w}$. Combining the both rates (\ref{Eq:decay-rate-high-T}) and (\ref{Eq:Samokhin}), we find
\begin{equation}
\frac{1}{\tau(\w)}\sim
\begin{cases}
\frac{(3+K_0^2)T^{1/2} \w^{3/2}}{\sqrt{K_0}m v_0^2}, &\w \ll \operatorname{min}(T,\tilde{\w}),
\\
\\
\frac{K_0(3+K_0^2)^2T \w}{D_K m^2 v_0^3}, & \tilde{\w}\ll \w\ll T.
\end{cases}
\label{Eq:self-consistent}
\end{equation}
In Eq. (\ref{Eq:self-consistent}) we implicitly assume that $\omega$ satisfies the inequality (\ref{Eq:w-bar}) such that the ``quasi MBL'' effects discussed at the end of Sec. \ref{Sec:golden-rule} (and in more detail in Sec.~\ref{Sec:level-spacings} below) do not matter. 

It is worth noting that both lines of the result (\ref{Eq:self-consistent})  can be obtained from a single self-consistent calculation.
Specifically, let us introduce self-consistency in  Eq.~\eqref{Eq:decay-rate-high-T} by replacing  $\xi(\w)$ (that essentially represents the disorder broadening of the plasmonic levels)  by $(1/\xi+1/v_0\tau)^{-1}$. Solving the emerging  equation for~$1/\tau$, one recovers Eq.~(\ref{Eq:self-consistent}).


\subsection{``High'' frequencies: ``Quasi MBL'' regime\label{Sec:level-spacings}}

The golden rule used in Sec.~\ref{Sec:golden-rule} is only applicable if the two-particle level spacing is smaller than the decay rate. In this section, we estimate the single-particle as well as the two-particle level spacing (cf. a related calculation in Supplemental Material of Ref.~\onlinecite{BanerjeeAltman16}). This allows us to establish the inequality (\ref{Eq:w-bar}) marking the onset of the ``quasi MBL'' regime where  the relaxation of high-energy plasmons (those with $\omega> \bar{\omega}$) is governed by the interaction with a bath of low-energy excitations with $\omega<\bar{\omega}$. We then analyze the inelastic decay of those high-energy plasmons.  

The number of plasmon modes per unit  energy  interval that have a finite overlap with a localized mode with energy $\w$ at position $x_1$ is given by
\begin{align}
N_1(\w) &\sim \int \diff x_2 \int \frac{\diff q_2}{2\pi}\e^{-|x_1-x_2|/\xi(\w_2)}\delta(\w-v_0|q_2|)
\\
&= \frac{\xi(\w)}{\pi v_0}.
\end{align}
The single-particle level spacing in the localization volume is thus given by
\begin{equation}
\Delta_1(\w)\sim \frac{1}{N_1(\w)}\sim \frac{v_0}{\xi(\w)},
\end{equation}
in agreement with the flat density of plasmonic modes. 

For the two-particle level spacing, we consider the number of different plasmon pairs 
that can emerge in a one-into-two decay of a plasmon with frequency $\omega$:
\begin{align}
\begin{split}
N_2(\w)& \sim \int \diff x_2 \int \diff x_3 \int\frac{\diff q_2}{2\pi}\frac{\diff q_3}{2\pi} \e^{-|x_1-x_2|/\xi(\w_2)}
\\
&\hspace{1.4cm}\times \e^{-|x_1-x_3|/\xi(\w_3)}\delta(\w-\w_2-\w_3)
\end{split}
\\
&=\int_{0}^{\infty} \frac{\diff \w_2}{\pi v_0}\int_{0}^{\infty} \frac{\diff \w_3}{\pi v_0} \xi(\w_2)\xi(\w_3)\delta(\w-\w_2-\w_3).
\end{align}
The energy  integrations should be cut on the lower limit at the single-particle level spacing $\Delta_1$ in order to cure the formal divergence.  Evaluating the integral over $\omega_3$ with the help of  the $\delta$-function and using Eq.~(\ref{Eq:localization-length}), we obtain
\begin{align}
N_2(\w)& \sim \int_{\Delta_1(\w)}^{\w-\Delta_1(\w)}\diff \w_2 \frac{\w_{\ast}^2}{(2\pi)^2}\frac{1}{\w_2^2(\w-\w_2)^2}
\\
&=\frac{\w_{\ast}^2}{2\pi^2}\int_{\Delta_1}^{\w/2}\diff \w_2 \frac{1}{\w_2^2(\w-\w_2)^2},
\end{align}
where $\w_{\ast}$ is given by Eq. (\ref{Eq:w_star}). 
The integral is dominated by the lower limit, yielding
\begin{equation}
N_2(\w)\sim\frac{\w_{\ast}^2}{\w^2\Delta_1(\w)}\sim\frac{\w_{\ast}^3}{\w^4}.
\label{Eq:N2}
\end{equation}

Using Eq.~(\ref{Eq:N2}), we find the corresponding two-particle level spacing:
\begin{equation}
\Delta_2(\w)\sim \frac{1}{N_2(\w)}\sim \frac{\w^4}{\w_{\ast}^3}\sim \frac{D_K^3\w^4}{v_0^3 K_0^6}.
\label{Eq:Delta2Final}
\end{equation}
Comparison of Eq. (\ref{Eq:Delta2Final}) to the golden-rule relaxation rate in the frequency range $\tilde{\w}\ll\w\ll T$ where the disorder is important [second line in Eq.~(\ref{Eq:self-consistent})] gives the criterion (\ref{Eq:w-bar}). 


We conclude that at low enough temperatures, 
\begin{equation}
T<T_0\equiv\frac{K_0^{7/2}(3+K_0^2)}{m D_K^2},
\label{Eq:T0}
\end{equation}
the relaxation of all the sub-thermal plasmons is correctly described by Eq. (\ref{Eq:self-consistent}). 
On the other hand, for $T>T_0$, the relaxation of high-frequency modes with $\w>\bar{\w}$ is different: it is governed by scattering off low-frequency excitations. 
It is clear however that the analysis leading to Eq. (\ref{Eq:decay-rate-high-T}) (see appendix \ref{App:decay-rate}) can be straightforwardly adapted to the present situation. Specifically, treating the scattering processes shown in Fig. \ref{Fig:processes} in the golden-rule manner, we should restrict the possible final states to those that contain a low-energy plasmon.  Proceeding along this direction, we find that in the frequency range
\begin{equation}
\tilde{\w}<\bar{\w}<\w<T,
\end{equation}
the plasmonic decay is described by
\begin{equation}
\frac{1}{\tau(\w)}\sim \frac{K_0(3+K_0^2)^2T \bar{\w}}{D_K m^2 v_0^3}, \quad \tilde{\w}<\bar{\w}<\w<T.
\label{Eq:rate-inefficient}
\end{equation}
Thus, the linear growth of the decay rate seen in the second line of Eq. (\ref{Eq:self-consistent}) saturates at high frequencies due to the ``quasi MBL'' effect.  Still, the rate is finite and no genuine  MBL of plasmons occurs.  


\subsection{Sub-thermal plasmons: Decay of plasmonic states  and plasmon diffusion\label{Sec:results}}

In Secs.  \ref{Sec:golden-rule}, \ref{Sec:self-consistency} and \ref{Sec:level-spacings} we have presented a detailed analysis of the inelastic relaxation of plasmons in a disordered LL.  We now summarize the main results obtained in these sections, focusing on the sub-thermal plasmons. 
We then establish a qualitative picture of plasmon dynamics that will allow us to analyze the thermal transport in the system in Sec. \ref{Sec:thermal_conductance}. 

\begin{figure}
\center
\includegraphics[scale=0.3]{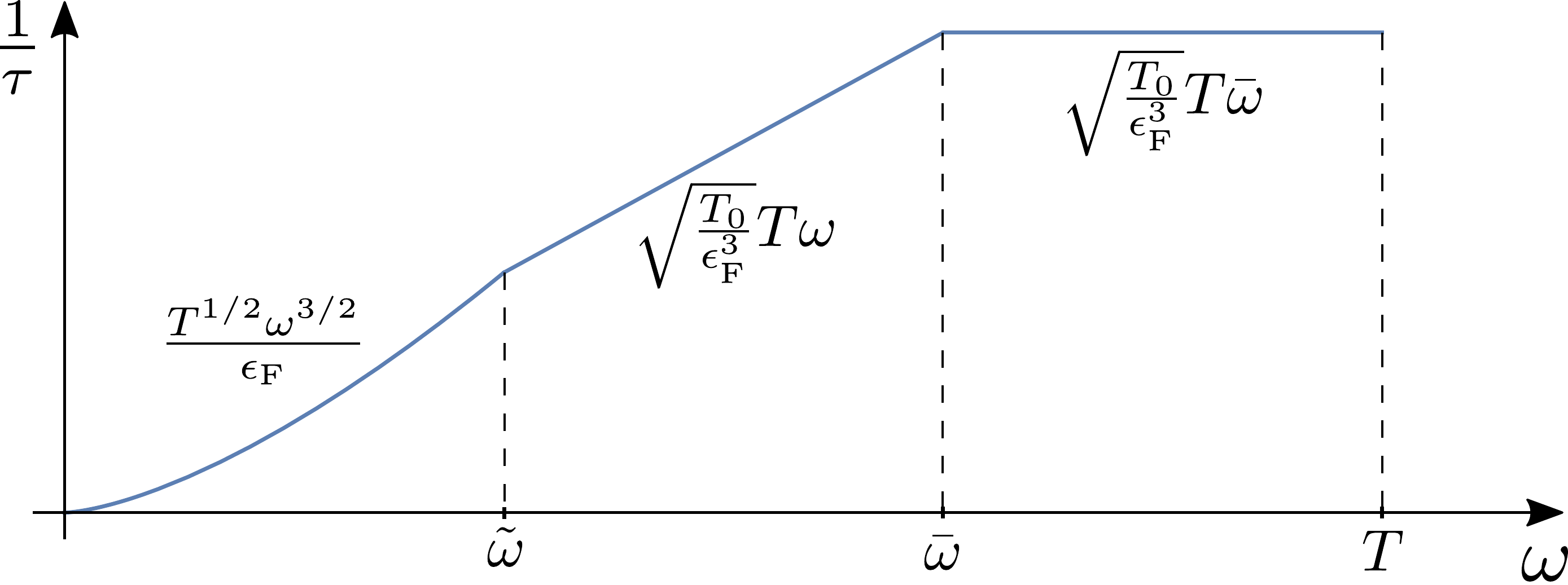}
\caption{Schematic behavior  of the inelastic scattering  rate  as a function of frequency, see Eqs. (\ref{Eq:self-consistent}) and (\ref{Eq:rate-inefficient}).  We omit the dependence of $\tau(\omega)$ on the interaction constant $K_0$ (assuming interaction strength of order unity). The characteristic frequency scales are $\tilde{\w}\sim T_0 T/\ef$ and $\bar{\w} \sim (TT_0^2)^{1/3}$. 
\label{Fig:decay-rate}}
\end{figure}

 Figure \ref{Fig:decay-rate} shows the schematic behavior  of the inelastic scattering  rate as a function of frequency for $\omega < T$. Our results for the decay rate are expressed there in terms of the characteristic energy scales $T_0$, $\tilde{\w}$, $\bar{\w}$ [see Eqs. (\ref{Eq:T0}),  (\ref{Eq:w_tilde}) and (\ref{Eq:w-bar})], and the { high-energy scale (``Fermi energy'')} $\epsilon_{\mathrm{F}}= mv_0^2/2 \gg T$. At high frequencies, $\omega>\bar{\omega}$, the plasmons are in the ``quasi MBL'' regime.  Interaction with low-energy plasmons is necessary for relaxation and the relaxation rate does not depend on frequency, see Eq.~(\ref{Eq:rate-inefficient}). The ``quasi MBL'' regime is pushed above temperature ($\bar{\omega}> T$) at low enough temperatures, $T<T_0$.

At lower frequencies, $\omega<\bar{\omega}$ (or, if $T<T_0$, for all sub-thermal bosons), the relaxation rate is given by Eq. (\ref{Eq:self-consistent}).  
Here, two regimes can be distinguished. At intermediate frequencies, $\tilde{\omega}<\omega<\bar{\omega}$, the relaxation of plasmons is appropriately described in terms of the decay of localized plasmonic modes. At lowest frequencies, $\omega<\tilde{\omega}$, one can neglect the disorder in the calculation of the inelastic lifetime, viewing these plasmons just as plane waves. (We will have, however, to recall about the presence of disorder when analyzing the contribution of these plasmons to thermal transport; see the discussion around Eq.~(\ref{Eq:diffusion-coefficient-low}).  

In Sec. \ref{Sec:thermal_conductance} we will analyze the thermal conductance in its dependence on the system size. For that purpose it is useful to discuss first the mechanism  of the bosonic transport. 

At frequencies larger than  $\tilde{\w}$, disorder leads to localization of plasmons if the anharmonicity  $\Ham_1$ is neglected. The relaxation induced by $\Ham_1$ goes via decay of a plasmon into two plasmons of similar energy (or recombination of two such plasmons). Typically, new plasmons are located a distance $\sim \xi(\omega)$, Eq. (\ref{Eq:localization-length}), apart from the location of the original plasmons, and the characteristic  time for such a decay process is $\tau(\omega)$ as given by Eqs.~(\ref{Eq:self-consistent}) and (\ref{Eq:rate-inefficient}). We can thus (somewhat loosely) think about the plasmon motion  as hops between the localized states and characterize this dynamics by a (frequency-dependent) effective ``diffusion coefficient'' $D_{\rm eff}(\omega)$ estimated as 
\begin{equation}
D_{\rm eff}(\w)\sim \frac{\xi^2(\w)}{\tau(\w)}\sim
\begin{cases}
\frac{v_0^2 T_0^{3/2} T}{\ef^{1/2}\w^3}, & \tilde{\w}<\w<\bar{\w},
\\
\\
\frac{v_0^2 T_0^{3/2}\bar{\w} T}{\ef^{1/2}\w^4}, & \bar{\w}<\w<T.
\end{cases}
\label{Eq:diffusion-coefficient-high}
\end{equation}
{ In Eq.~\eqref{Eq:diffusion-coefficient-high} and in the sequel, we omit the dependence on the interaction constant $K_0$ (assuming
interaction strength of order unity).}
It is worth noting that the energy of plasmons with given frequency is not a conserved quantity, so that $D_{\rm eff}(\omega)$ is not a diffusion coefficient in a strict sense. Rather, it is an effective quantity that characterizes the dynamics of plasmons with frequency $\sim \omega$ at not too long time scales and thus their contribution to the thermal transport. 

Let us now discuss the dynamics of plasmons at lowest energies, $\omega<\tilde{\omega}$. Here plasmons are plane waves with velocity $v_0$ that experience inelastic collisions (decay or recombination) within the time $\tau(\omega)$ given by the first line in Eq. (\ref{Eq:self-consistent}). The corresponding mean-free path is $l^{\mathrm{in}}(\omega)=v_0 \tau(\omega)$.  As in the case of higher frequencies, we ascribe to these low-frequency plasmons an effective diffusion coefficient $D_{\rm eff}(\omega)$,
\begin{equation}
D_{\rm eff}(\omega) \sim v_0 l^{\mathrm{in}}(\w) = v_0^2 \tau(\omega) \sim \frac{v_0^2 \,\ef}{T^{1/2}\w^{3/2}}, \qquad \w<\tilde{\w}.
\label{Eq:diffusion-coefficient-low}
\end{equation}
While Eq.~(\ref{Eq:diffusion-coefficient-low}) looks pretty standard at first sight, it is not entirely trivial. Indeed, the scattering processes that we considered while calculating the inelastic time $\tau(\omega)$ conserve the total momentum and thus the energy current. They are thus not sufficient to establish the heat diffusion, since the latter requires some mechanism of momentum relaxation. In a {\it clean} LL the relaxation of the total plasmon momentum is established via the plasmon Umklapp scattering\cite{SamantaEtAl19,MatveevRistivojevic19} that in the fermionic description of the LL corresponds to the equilibration of the number of left- and right-moving fermions. 
The latter is possible only via the diffusion of a deep hole in a Fermi sea through the bottom of the energy band and is thus exponentially suppressed at low temperatures. 
In the present case of a  {\it disordered}  LL there is a much more efficient mechanism of relaxing the bosonic momentum: the back-scattering of a plasmon by randomness. The crucial role of this mechanism is obvious in Eq.~(\ref{Eq:diffusion-coefficient-high}) which describes a hopping-like relaxation of localized states. Correspondingly, the strength of the disorder enters explicitly Eq.~(\ref{Eq:diffusion-coefficient-high})  via $T_0$ and $\bar{\omega}$. But where is the disorder in the derivation of  Eq.~(\ref{Eq:diffusion-coefficient-low})? The answer is as follows. Considering the evolution of an initial low frequency plasmon with $\omega< \tilde{\omega}$, we make two important observation: 
 (i) in an inelastic scattering process contributing to $\tau(\omega)$, the plasmons that are created or annihilated can typically have energies that are larger than $\omega$ by a factor $\sim 2$; (ii) the rate of inelastic scattering grows with frequency as $\omega^{3/2}$. It follows that the energy of the $\omega$ plasmon is transported to higher  frequencies $\sim \tilde{\omega}$ within a time $\sim \tau(\omega)$. Since the disorder-induced backscattering time $\xi (\tilde{\omega}) / v_0$ is equal to $\tau (\tilde{\omega})$ and thus is much shorter than $\tau(\omega)$, the overall ``transport'' scattering time at which the relaxation of the momentum of the $\omega$ plasmon takes place is $\sim \tau(\omega)$. This provides a justification to Eq.~(\ref{Eq:diffusion-coefficient-low}). It is worth stressing that, in this argument, we have used the condition $\tilde{\omega} < T$, which implies that the disorder is not too weak. 
 
\begin{figure}
\centering
\includegraphics[scale=0.3]{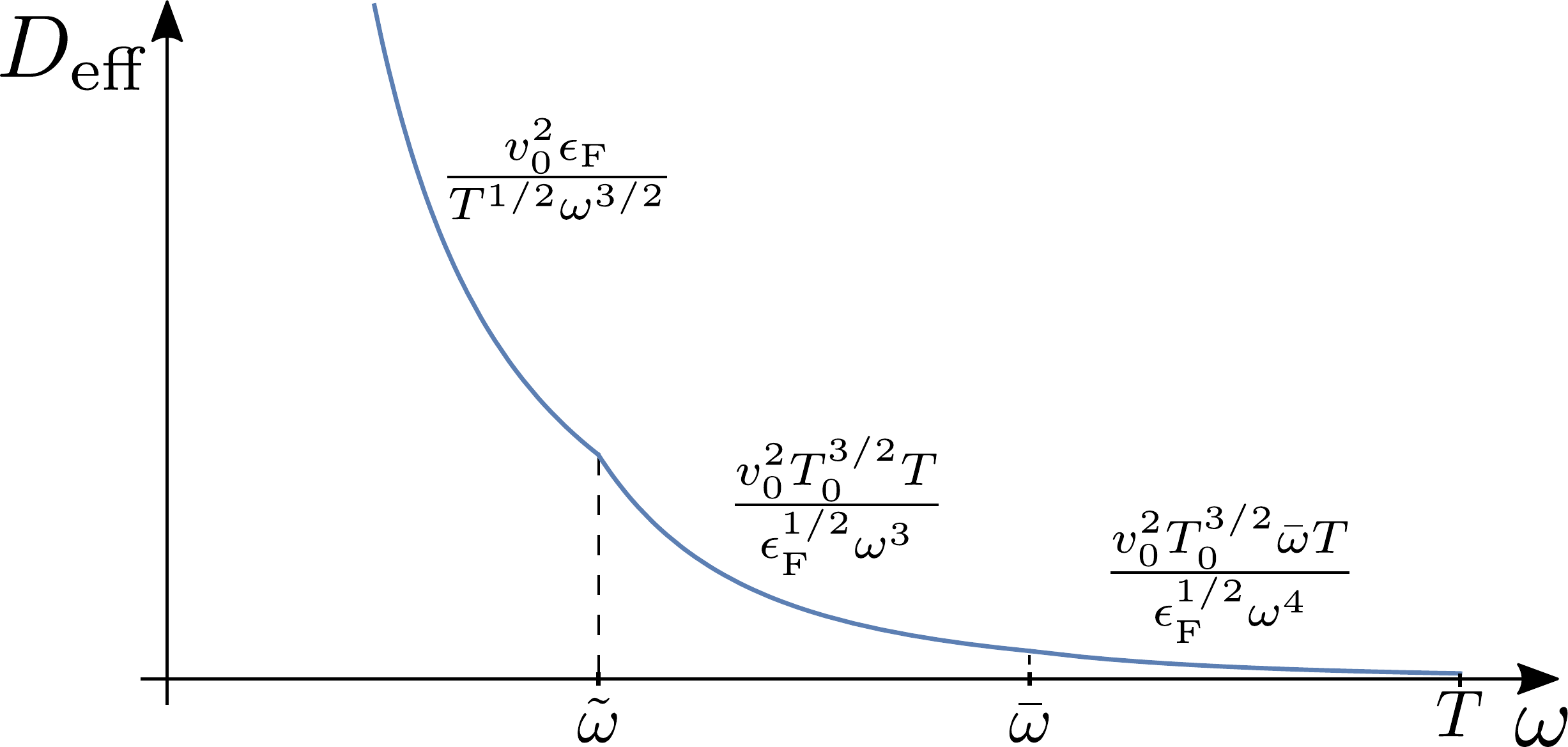}
\caption{Schematic behavior of the effective diffusion coefficient  as predicted by  Eqs.~\eqref{Eq:diffusion-coefficient-low} and \eqref{Eq:diffusion-coefficient-high}
We omit the dependence of $D_{\rm eff}(\omega)$ on the interaction constant $K_0$ (assuming interaction strength of order unity). The characteristic frequency scales are $\tilde{\w}\sim T_0 T/\ef$ and $\bar{\w} \sim (TT_0^2)^{1/3}$. 
\label{Fig:diffusion-coefficient}}
\end{figure}

The behavior of the effective diffusion coefficient as predicted by Eqs. (\ref{Eq:diffusion-coefficient-high}) and  (\ref{Eq:diffusion-coefficient-low}) is summarized in Fig.~\ref{Fig:diffusion-coefficient}. 
In the next section we will use these results to  explore the dependence of the thermal conductance of the disordered LL on the length of the system.

\section{Thermal conductance \label{Sec:thermal_conductance}}

We are now in a position to study the thermal transport in a disordered LL. Let us consider a system of length $L$ connected to two reservoirs with slightly different temperatures, $T_R-T_L=\Delta T$. In the linear-response regime that we are considering, the thermal current $j_E$ is proportional to the temperature difference,
\begin{equation}
j_E = G \Delta T,
 \end{equation}
where $G$ is the thermal conductance. Our goal in this section is to determine the behavior of $G$ and, in particular, its dependence on $L$. 

The thermal current induced by the temperature gradient can be divided into two parts. 
First, since the scattering processes become progressively weaker with lowering frequency, plasmons with a sufficiently low energy,
\begin{equation}
 \omega<\omega_c(L),
 \end{equation}
 traverse the whole system ballistically. Here, $\omega_c(L)$ is the cutoff frequency for the ballistic motion that will be determined below. 
 The energy current carried by the ballistic plasmons is given by  
 \begin{equation}
j_{E}^{\mathrm{bal}}=2\int_{0}^{\omega_c}\frac{\diff \omega}{2\pi}  \omega \left[ n_L(\w)-n_R(\w) \right],
\end{equation}
where $n_R$ and $n_L$ denote the Bose distributions in the reservoirs.  Using the low-frequency asymptotics  of the Bose distribution, $n_i(\omega) \simeq T_i/\omega$, we find that the corresponding contribution to the thermal conductance is given, up to a numerical coefficient, by the cutoff frequency:
\begin{equation}
G^{\mathrm{bal}}(L)=\frac{\w_c(L)}{\pi}.
\label{Eq:GBal}
\end{equation}
It remains to understand the cutoff for the ballistic motion.  For comparatively short systems, $L< \xi(\tilde{\omega})$, the process that cuts off the ballistic propagation is the elastic scattering, so that $\omega_c(L)$ is defined by the equation
\begin{equation}
\xi(\omega_c)=L\,, \qquad L<\xi(\tilde{\omega}).
\end{equation} 
On the other hand, in a longer system the ballistic contribution is cut off by the inelastic length $l^{\mathrm{in}}(\omega)=v_0 \tau(\omega)$ with $\tau$ given by the first line of Eq.(\ref{Eq:self-consistent}). This length limits the ballistic motion of a low-frequency plasmon due to a combined effect of the anharmonicity and disorder,
see Eq.~(\ref{Eq:diffusion-coefficient-low}) and discussion after it. We thus get the condition on $\omega_c(L)$ for long systems:
\begin{equation}
v_0 \tau(\omega_c)=L\,, \qquad L > \xi(\tilde{\omega}).
\end{equation} 
Employing Eqs. (\ref{Eq:localization-length}), (\ref{Eq:self-consistent}), and (\ref{Eq:GBal}), we finally obtain the contribution of ballistic plasmons to the thermal conductance:
\begin{equation}
G^{\mathrm{bal}}(L)\sim
\begin{cases}
\displaystyle \frac{T_0^{1/4} \ef^{1/4} v_0^{1/2}}{L^{1/2}}, \quad L<\xi(\tilde{\w}),
\\[0.5cm]
\displaystyle \frac{\ef^{2/3}v_0^{2/3}}{T^{1/3}L^{2/3}}, \qquad \ L>\xi(\tilde{\w}).
\end{cases}
\label{Eq:GBalFinal}
\end{equation}

The bosons with frequency larger than $\omega_c(L)$ provide the second contribution to the thermal transport. As they travel diffusively, their contribution is characterized by a thermal  {\it conductivity} $\kappa$ that can be related to the effective diffusion coefficient $D_{\rm eff}(\omega)$, Eqs.~(\ref{Eq:diffusion-coefficient-high}) and (\ref{Eq:diffusion-coefficient-low}),
\begin{equation}
\kappa(L)\simeq T\int_{\omega_c(L)}^T d\omega \: D_{\rm eff}(\omega).
\label{Eq:kappa}
\end{equation}
The corresponding contribution to the thermal conductance of the sample is
\begin{equation}
\delta G(L)=\frac{\kappa(L)}{L}.
\end{equation} 
It is now easy to verify that the integration in Eq. (\ref{Eq:kappa}) is dominated by the lower limit and reproduces the ballistic contribution (\ref{Eq:GBalFinal}). The full thermal conductance $G(L) = G^{\mathrm{bal}}(L) + \delta G(L)$  of the system is thus correctly described by the ballistic result, Eq. (\ref{Eq:GBalFinal}), governed by plasmons with frequencies $\omega \sim \omega_c(L)$ (i.e., those on the upper border of the ballistic range of frequencies).  The plasmons with $\omega \gg \omega_c(L)$ contributing to $\delta G$ yield in addition only sub-leading  (in powers of $1/L$) terms. 

We thus conclude that the dependence of the thermal conductance of a disordered [in the sense of $K(x)$] nonlinear LL on the size of the system is given by Eq. (\ref{Eq:GBalFinal}). In   comparatively short samples, it is determined solely by the elastic scattering and our result reproduces the $L^{-1/2}$ scaling found previously\cite{KhmelnitskiiEtAl98,Amir_Oreg_Imry18, Dhar01} within the harmonic approximation.  
In longer systems, $L>\xi(\tilde{\omega})$, the anharmonicity (or, equivalently, the plasmon interaction) plays a crucial role. We find that $G(L)$ scales in this regime as   $G(L)\propto L^{-2/3}$. The same scaling, known from studies of clean  classical  systems at high temperature \cite{Narayan02,Spohn14}, was recently predicted to occur in a clean LL\cite{SamantaEtAl19} but only for exponentially long system sizes (larger than the plasmon Umklapp mean free path).



\section{Summary and discussion\label{Sec:Summary}}

We have studied the relaxation of plasmonic modes and thermal transport {in a ``$K$-disordered'' nonlinear LL, i.e., in a LL with spatially  varying interaction parameter $K$ and cubic anharmonicity.} 
Within the harmonic approximation, the disorder in $K$ leads to localization of plasmons with the localization length $\xi\propto 1/\omega^2$, in agreement  with previous findings. 
  
The anharmonicity  leads to the  interaction among plasmons, which induces relaxation of plasmonic modes. Figure \ref{Fig:decay-rate} summarizes our results for the plasmonic decay rate. At low frequencies,  $\omega < \tilde{\w}$, the localization length is longer than the inelastic mean free path. 
The resulting scaling of the relaxation rate $1/\tau\propto \omega^{3/2}$ coincides with the corresponding result for a clean LL\cite{Samokhin98}. 
At higher frequencies the disorder effects set in, and the rate scales linearly, $1/\tau\propto \omega$,  until saturation above the frequency $\bar{\omega}$ 
[given by Eq.~(\ref{Eq:w-bar})] where the plasmons enter the ``quasi MBL'' regime.  Such ``high-energy'' ($\omega \gg \bar{\omega}$) plasmons can not relax via interactions with plasmons of similar frequency due to MBL-like physics. The true MBL can not develop, however, because of the interaction with the plasmonic bath at low energies. This mechanism of the destruction of MBL bears a certain similarity to the one known\cite{GMMP2017} for continuous disordered fermionic models where the unbounded growth of the single-particle localization length at high energies paves the way for relaxation.  An important difference in our case in comparison to the fermionic models is the presence of the bath at {\it low} energies. The relaxation assisted by such a bath does not require exponentially rare events and is thus suppressed only in a power-law fashion in comparison to the naive golden-rule result.   
 
We have analyzed the plasmon dynamics and demonstrated that it can be  characterized by an effective diffusion coefficient $D_{\rm eff}(\omega)$  whose frequency dependence is summarized in Fig.~\ref{Fig:diffusion-coefficient}.  At low frequencies, $\w<\tilde{\w}$, where $\tilde{\w}$ is given by Eq.~(\ref{Eq:w_tilde}),
the fastest scattering mechanism is the inelastic scattering that transports the plasmon energy flux towards higher frequencies where it is backscattered by disorder. As a result, $D_{\rm eff}(\omega)$ is determined  by the inelastic scattering rate.
 At higher frequencies, $\w>\tilde{\w}$ the plasmon dynamics can be visualized as interaction-induced hops between localized states. 
  
The thermal conductance $G(L)$ of the disordered LL shows a non-trivial scaling with  the size of the system $L$, see Eq.~(\ref{Eq:GBalFinal}).  The conductance $G(L)$ is dominated by the low-frequency plasmons that are ballistic on the scale of the system size.  Upon increase of the size of the system, the scaling of $G(L)$ crosses over from $G(L)\propto L^{-1/2}$ (effectively noninteracting plasmons) to $G(L)\propto L^{-2/3}$. The latter result does not contain the disorder explicitly as it derives from the inelastic scattering length in a clean system. However, the disorder enters implicitly through the condition $\tilde{\w} \ll T$, which implies that the disorder is not too weak.  

Before closing this paper, let us discuss possible extensions of our work. First, 
as was pointed out in Sec.~\ref{Sec:Model},
we have limited our consideration to the case of not too strong, uncorrelated disorder. The strong-disorder effects  and/or the correlated nature of disorder may alter significantly\cite{Gurarie2002, Gurarie2003, Refael2013,  Amir_Oreg_Imry18} the density of states and the localization properties of the single-particle wave functions of the low-energy plasmons. Implications of the anharmonic coupling of plasmons in such systems constitute an interesting direction for future research. 

Second,  one can include in the model the fermionic backscattering (or, equivalently, QPS in the context of the JJ chains). We remind the reader that we have not included such terms in view of their exponential smallness, see Sec.~\ref{Sec:Intro}. On the other hand, the fermionic backscattering in a LL grows under the renormalization group for a repulsive interaction (or a not too strong attraction). In this situation, it will modify the ultimate infrared behavior of the system. 

A further ingredient that was discarded in our analysis is the curvature of the plasmonic spectrum. While it is not expected to affect the low-frequency (and thus large-$L$) behavior, it may  influence significantly\cite{SamantaEtAl19} the  inelastic relaxation of plasmons at intermediate frequencies, cf. Ref.~\onlinecite{SamantaEtAl19}.
In the context of JJ chains, both the curvature of plasmons and the QPS are of direct relevance to spectroscopy of real physical systems\cite{ManucharyanEtAl18}.

Finally,  and on a more general note, it would be interesting to explore possible connections between the model of  the $K$-disordered  nonlinear LL and problems of  the generalized 1D hydrodynamics\cite{Castro2016,Bertini2016,Caux2017,Bulchandani2017,Bulchandani2018,Nardis2019,Gopalakrishnan2019,Agrawal2019} that currently attract much attention.



\begin{acknowledgments}

We thank D. B. Gutman and R. Samanta for useful discussions. 
IP acknowledges support by the Swiss National Science Foundation. 

 \end{acknowledgments}



\appendix

\section{Relation to model of fluctuating electron density\label{App:comparision-Khmelnitskii}}

In this appendix, we give a brief account of the derivation of the Hamiltonian (\ref{Eq:Hamiltonian}) in the context of a quantum wire with smooth disorder potential\cite{KhmelnitskiiEtAl98,GramadaRaikh97} and establish the correspondence of our notations to those of  Ref.~\onlinecite{KhmelnitskiiEtAl98}.


The Hamiltonian  describing an inhomogeneous electron liquid in 1D is given by\cite{KhmelnitskiiEtAl98}
\begin{equation}
\Ham^{\mathrm{FHK}}=\Ham_0^{\mathrm{FHK}}+\Ham_1^{\mathrm{FHK}},
\end{equation}
where the quadratic Hamiltonian can be written as $\Ham_0^{\mathrm{FHK}}=\int \diff x H_0^{\mathrm{FHK}}(x)$ with
\begin{equation}
H_0^{\mathrm{FHK}}(x)=\frac{p^2}{2m n(x)}+\frac{1}{2}\left(V_0+\frac{\pi^2}{m}n(x)\right)\left(\frac{\diff [n u]}{\diff x}\right)^2.
\end{equation}
Here, $V_0$ denotes the zero momentum component of the interaction potential for fermions, and $n(x)=n_0+\delta n(x)$ denotes the local average electron density which is assumed to fluctuate randomly around the mean value $n_0=\pf/\pi$, where $\pf$ is the Fermi momentum. The length scale of the fluctuations of the disorder potential leading to the fluctuating density is assumed to be much larger than the Fermi wavelength such that backscattering of fermions is not important. The fluctuations of the density are assumed to be short-range correlated,
\begin{equation}
\overline{\delta n(x)\delta n(\xp)}=n_0^2 l_D \delta(x-\xp).
\end{equation}
The displacement field $u(x)$ and the momentum $p(x)$ obey canonical commutation relations and are related to the operators $\phi$ and $\theta$ of our model via
\begin{equation}
\frac{1}{\pi}\phi(x)=n(x)u(x), \qquad \partial_x\theta(x)=\frac{1}{n(x)}p(x).
\label{Eq:correspondences_operators}
\end{equation}
Comparing the Gaussian model of Ref.~\onlinecite{KhmelnitskiiEtAl98} to our theory, we obtain the following correspondences:
\begin{equation}
\begin{split}
v(x)&=\frac{1}{m}\sqrt{n(x)[\pi^2 n(x)+m V_0]}, 
\\
K(x)&=\frac{\pi \sqrt{n(x)}}{\sqrt{\pi^2 n(x)+mV_0}},
\\
K_0&=\frac{\pi\sqrt{n_0}}{\sqrt{\pi^2 n_0 +m V_0}}=g, 
\\
D_K&=\frac{K_0^2}{4}(1-K_0^2)^2l_D=\frac{g^6 V_0^2 l_D}{4\pi^2 \vf^2}.
\end{split}
\end{equation}
Here, $\vf$ denotes the Fermi velocity. 

The nonlinear term $\Ham_1^{\mathrm{FHK}}$ is identical to the term $\Ham_1$ in Eq.~\eqref{Eq:H1} as can be seen using the correspondences \eqref{Eq:correspondences_operators}. \footnote{Equation (11) of Ref.  ~\onlinecite{KhmelnitskiiEtAl98} contains a typo: a missing   factor of  $1/n(x)$ in the first term.}



\section{Calculation of the localization length \label{App:localization-length}}

In this appendix, we give a brief account of the analysis of the localization  length of non-interacting plasmons, Sec.~\ref{Sec:Localization_Length}. 

We expand the plasmonic fields $\phi$ and $\theta$ into normal modes according to\cite{GiamarchiBook}
\begin{equation}
\begin{split}
\phi(x)&=-\frac{i\pi}{L}\sum_{q\neq 0}\left(\frac{K_0 L|q|}{2\pi}\right)^{1/2}\frac{1}{q}\e^{-i q x} \left(b_q^{\dagger}+b_{-q}\right),
\\
\theta(x)&=\frac{i\pi}{L}\sum_{q\neq 0}\left(\frac{L|q|}{2\pi K_0}\right)^{1/2}\frac{1}{|q|}\e^{-i q x} \left(b_q^{\dagger}-b_{-q}\right),
\end{split}
\label{Eq:relation-to-bosonic-operators}
\end{equation}
where $L$ is the system size. The homogeneous part of the Hamiltonian (\ref{Eq:H0}) takes the diagonal form
\begin{equation}
\Ham_0^{\mathrm{LL}}=\sum_q v_0 |q| b_{q}^{\dagger}b_q+\mathrm{const.}
\end{equation}

The rate of scattering  induced by the random fluctuations $\delta v(x)$ and $\delta K(x)$ can be extracted from the collision integral,
\begin{equation}
I[f]=-\sum_{q^{\prime}}W_{qq^{\prime}}[f(q)-f(q^{\prime})],
\label{Eq:collision-integral-disorder}
\end{equation}
with the transition probability given by the golden rule
\begin{equation}
W_{qq^{\prime}}=2\pi \, \overline{|\langle 0 | b_{q^{\prime}} \Ham_0^{\mathrm{dis}} b_q^{\dagger}|0\rangle|^2}\,\delta(v_0 |q|-v_0|q^{\prime}|).
\end{equation}
Here $|0\rangle$ is the empty state, $f(q)$ is the plasmonic distribution function, and $\Ham_0^{\mathrm{dis}}$ is the random part of the Hamiltonian,
Eq.~(\ref{Eq:perturbation}). 

After the disorder averaging, the modulus square of the matrix element takes the form
\begin{equation}
\begin{split}
\overline{|\langle 0 | b_{q^{\prime}} \Ham_0^{\mathrm{dis}} b_q^{\dagger}|0\rangle|^2}=\frac{1}{4L}|&qq^{\prime}|\Bigl\{D_v[1+\sign(qq^{\prime})]^2
\\
&+\frac{v_0^2}{K_0^2}D_K [1-\sign(qq^{\prime})]^2\Bigr\}.
\end{split}
\end{equation}
Note that the correlations between $v(x)$ and $K(x)$ do not contribute. Moreover, due to the structure of the collision integral, Eq.~\eqref{Eq:collision-integral-disorder}, the fluctuations of the velocity contribute only to forward scattering and drop out of the kinetic equation. As a result, the kinetic equation describing the evolution of the distribution function of the plasmons in the presence of disorder in the parameters of the quadratic Hamiltonian $\Ham_0$ is given by
\begin{equation}
\frac{\partial f(q)}{\partial t}=-\frac{v_0 D_K}{K_0^2}q^2[f(q)-f(-q)].
\end{equation}
From this equation, we extract the elastic scattering time $\tau_{\mathrm{el}}(\w)$ which leads to the localization length (mean free path) of the plasmons, 
\begin{equation}
\xi(\w)=v_0 \tau_{\mathrm{el}}(\w)=\frac{v_0^2K_0^2}{2 D_K \w^2},
\end{equation}
which is Eq.~\eqref{Eq:localization-length} of the main text.



\section{Details of the perturbative calculation of the decay rate \label{App:decay-rate}}

In this appendix, we present details of the golden-rule calculations of the plasmon decay rate leading to Eqs.~\eqref{Eq:decay-rate-high-T}  and  \eqref{Eq:decay-rate-low-T}.

We begin by explicitly computing the matrix element \eqref{Eq:definition-matrix-element}. Making use of the transformation \eqref{Eq:transformation-to-bosons}, we express the nonlinearity \eqref{Eq:H1} in terms of bosonic operators. [Here we can set $v(x)=v_0$ and $K(x)=K_0$ in Eq.~\eqref{Eq:transformation-to-bosons}]. The evaluation of the vacuum expectation value in Eq.~\eqref{Eq:definition-matrix-element} yields 
\begin{equation}
\begin{split}
\langle 0|b_{\mu_3}b_{\mu_2}\Ham_1 b_{\mu_1}^{\dagger}|0\rangle \simeq & \frac{i \sqrt{\pi}}{2\sqrt{2}m\sqrt{K_0}}(3+K_0^2)\sqrt{|q_1q_2q_3|}
\\
& \times \int \diff x \, \psi_{\mu_1}(x)\psi_{\mu_2}^{\ast}(x)\psi_{\mu_3}^{\ast}(x).
\end{split}
\end{equation}
We have assumed here that the disorder is not too strong, in the sense of the condition \eqref{Eq:w_star}, i.e., the localization lengths of involved states are larger than the corresponding wave lengths,  $|q_i|\xi(\w_i)\gg 1$.  In view of this, we have the approximations $\psi_{\mu_i}^{\prime}(x) \simeq i q_i\, \psi_{\mu_i}(x)$ as well as $\Omega_{\mu_i}\simeq \w(q_i)$. Inserting this result for the matrix element into Eq.~\eqref{Eq:rate-initial} and performing the integrations over $x_2$ and $x_3$ yields
\begin{widetext}
\begin{equation}
\begin{split}
\frac{1}{\tau(q_1)}=\frac{\pi^2(3+K_0^2)^2}{4v_0K_0 m^2 L^2}&\sum_{q_2,q_3}|q_1q_2q_3|\delta(|q_1|-|q_2|-|q_3|)[1+\nB(\w_2)+\nB(\w_3)]\int \diff x \diff \xp \e^{i(q_1-q_2-q_3)(x-\xp)}
\\
&\times \left(1+\frac{|x-\xp|}{\xi_2}\right)\e^{-|x-\xp|/\xi_2}\left(1+\frac{|x-\xp|}{\xi_3}\right)\e^{-|x-\xp|/\xi_3}\frac{1}{\xi_1}\e^{-|x-x_1|/\xi_1}\e^{-|\xp-x_1|/\xi_1},
\end{split}
\label{Eq:rate-semi-final}
\end{equation}
where $\w_i=\w(q_i)=v_0|q_i|$ and $\xi_i=\xi(\w_i)$. After introducing relative ($r=x-\xp$) and center-of-mass [$R=(x+\xp)/2$] coordinates, we perform the integration over $R$ and estimate the integration over $r$ by cutting the integral at the upper limit at $\xi_1$. In the limit $L\to \infty$, the sums over $q_2$ and $q_3$ are replaced  by the integrations and we get
\begin{equation}
\frac{1}{\tau(q_1)}\sim \frac{(3+K_0^2)^2}{v_0K_0 m^2}\int \diff q_2 \int \diff q_3 |q_1 q_2 q_3|\,\frac{\sin[(q_1-q_2-q_3)\xi_1]}{q_1-q_2-q_3}\,\delta(|q_1|-|q_2|-|q_3|)[1+\nB(\w_2)+\nB(\w_3)].
\end{equation}
\end{widetext}
The integration over $q_3$ can be performed by exploiting the delta-function, the remaining integral over $q_2$ can be estimated in different limits. In the case of $\w_1=v_0|q_1|\ll T$, the Bose functions can be approximated by their low-frequency behavior, $n_B(\omega) \simeq T/\omega$. We obtain in this limit
\begin{equation}
\frac{1}{\tau(\w_1)}\sim\frac{(3+K_0^2)^2\xi(\w_1)}{K_0m^2 v_0^5}\w_1^3 T, \qquad \w_1 \ll T,
\end{equation}
which is Eq.~\eqref{Eq:decay-rate-high-T} of the main text.
In the opposite limit of high frequencies, we get 
\begin{equation}
\frac{1}{\tau(\w_1)}\sim\frac{(3+K_0^2)^2\xi(\w_1)}{K_0m^2 v_0^5}\w_1^4, \qquad \w_1 \gg T,
\end{equation}
which is Eq.~\eqref{Eq:decay-rate-low-T} of the main text.



\bibliography{references.bib}

\end{document}